\documentclass[conference, a4paper]{IEEEtran}

\ifCLASSINFOpdf
\else
\fi

\usepackage[utf8]{inputenc}

\usepackage{hyperref}

\usepackage{graphicx}
\usepackage{xcolor}
\usepackage{epstopdf}
\usepackage{amsfonts}

\usepackage[caption=false, font=footnotesize]{subfig}
\begin{document}

\title{Analyzing the Rework Time and Severity of Code Debt: A Case Study Using Technical Debt Catalogs}

\author{\IEEEauthorblockN{Bruno Santos de Lima,
Rogério Eduardo Garcia}
\IEEEauthorblockA{Department of Mathematics and Computation,\\São Paulo State University (UNESP), Faculty of Science and Technology\\Presidente Prudente, São Paulo, Brazil\\
E-mail: \{bruno.s.lima, rogerio.garcia\}@unesp.br}}

\maketitle

\begin{abstract}
This paper presents a case study analyzing Hibernate ecosystem software projects to investigate and demonstrate Code Debt behavior in relation to severity and rework time. The case study carried out revealed that the Code Debt with severity related to impact on software maintenance is the most representative and has the largest rework times to be paid in the Hibernate ecosystem. Besides, it was found that the distributions of rework times of Code Debt for all severities undergo variations in the initial versions of the development.
\end{abstract}

\begin{keywords}
Code Debt, Technical Debt, Technical Debt Analysis, Software Evolution, Mining Software Repositories
\end{keywords}

\IEEEpeerreviewmaketitle

\section{Introduction}
\label{sec:introduction}

\noindent 
In software development, several standards are followed aiming at achieving the best possible quality in the final product. In this context, Software Engineering is essential, however, it is also necessary to pay attention to how the software source code is implemented so that it is efficient and readable avoiding difficulties in the maintenance stage. For this, to follow technical activities and good programming practices is fundamental.

In the literature, the Technical Debt metaphor is presented describing the cost-benefit relationship between postponing of technical software development activities to obtain short-term profits and the long-term consequences~\cite{kruchten:2012}. The term was originally used with the focus on coding practices, the metaphor subsequently had its concept expanded including architecture, testing, and even software documentation~\cite{guoSeaman:2011}.

The presence of Technical Debt and their accumulation make the maintenance phase more complex and costly, besides degrading the quality of the software. The effective management Technical Debt is a fundamental activity to achieve and maintain the quality of the software~\cite{rios:2018}. Some cases are reported in which the inattention in the monitoring of Technical Debt, make the maintenance process infeasible in such a way that each new functionality added to the software will undermine existing implementations~\cite{seaman:2015}~\cite{nugroho:2011}. Thus, consider information about Technical Debt, analyzing its behavior, evolution, and impacts become important to subsidize its management.

In the source code perspective, Technical Debt is described as poorly written code that violates good practices or coding rules, denominates Code Debt~\cite{li:2015}~\cite{rios:2018}. In the literature are found works that analyze Technical Debt present in the source code from different perspectives, for example, an identity which problems are most frequent, which phases of development are more likely to accumulate these problems, or the same, which aspects have more influence for insertion of new debt instances.

However, each problem has a grade of impact and time required for its correction, these two characteristics being variant according to the type of problem. In this context, the objective of this paper is to analyze the behavior of Code Debt from the perspective of severity and time required for resolution, and the possible relations between these two aspects. To enable this analysis, we conducted a case study in which we identified and cataloged the Technical Debt present in the source code belonging to the Hibernate ecosystem software. To guide our analysis we define as research questions:

\textbf{RQ1:} What is the distribution of rework time for each severity group of Code Debt? Is this distribution similar in all analyzed software of this ecosystem?

\textbf{RQ2:} Is the distribution of rework time for each severity group experiencing significant variation across the multiple versions of each software of the Hibernate ecosystem?

\textbf{RQ3:} Analyzing the distribution of the total amount of Code Debt of each severity group, which groups are most representative in the Hibernate ecosystem?

\textbf{RQ4:} Observing the evolution of the total rework time of each severity group, is the Code Debt controlled in the Hibernate Ecosystem?

This paper is organized as follows: Section~\ref{sec:background} background is described. Section~\ref{sec:relacionados} the related works are presented. Section~\ref{sec:setup} outlines the case study setup followed for the accomplishment of this work. Section~\ref{sec:resultados} the results obtained after the study of the case are presented and discussed. Section~\ref{sec:conclusoes} presents the final remarks and future works.


\section{Background}
\label{sec:background}

The Technical Debt metaphor was evidenced in 1992 through an experience report, describing a situation where code quality was exchanged for general short-term gains~\cite{cunningham:1992}. The software development community uses the term Technical Debt to label internal software quality problems, future development, and maintenance risks, as well as the costs required to solve these problems which often come from improper shortcuts made by software engineers to bring business benefits~\cite{ramasubbu:2014}. Considering the financial point of view, an analogy with the financial debt can be perceived, because the Technical Debt also results in \textit{interest}, extra efforts, that need to be paid in the future during the software development process.

The term was originally used with a focus on coding practices, and then the metaphor had its concept expanded to include a requirement, architectural, test, defect, documentation, among others~\cite{li:2015}. Thus, a Technical Debt may be related to a specific type of artifact software~\cite{guoSeaman:2011}. In the perspective of source code, the Technical Debt is described as poorly written code, duplicates, with low readability, which infringe to some extent the good coding practices~\cite{li:2015}. Also, Code smells, classes or complex methods, violations of modularity and the principles of Object Orientation, are commonly called Code Debt~\cite{rios:2018}.

These problems can be identified through the Automatic Static Analysis (ASA) of source code. This type of analysis looks for violations of recommended programming practices that may cause failures or degrade software quality to some extent, this analysis is performed without the source code being executed~\cite{alfayez:2018}.


\section{Related Works}
\label{sec:relacionados}

In the literature are present studies that analyze from a different perspective the behavior of problems present in source code characterized as Code Debt. Tufano\textit{et al.}~\cite{tufano:2015} analyzed the changing history of a set of open-source software projects to understand why Code Smells were introduced and their life cycle. It has been identified that most Code Smells are introduced in the process of creating classes and files or implementing and enhancing functionality. In a complementary study, Jesus \textit{et al.}~\cite{jesus:2017} identified that the aspects: the size of the project, number of employees and speed of development are the most related with montant the Technical Debt present in the software.

Digkas \textit{et al.}~\cite{digkas:2017} analyzed the evolution of Code Debt present in the Apache ecosystem to investigate how the Code Debt evolves considering each type of problem identified. Was identified 232 different types of problems incurred in Technical Debt, using the SonarQube tool, among them were pointed out the most frequent and those that take longer to be corrected over time. Posteriorly, Digkas \textit{et al.}~\cite{digkas:2018} identified what types of problems get paid overtime. This work differs from those found in the literature, once, through a case study, aimed at analyzing the behavior of Code Debt regarding severity and rework time, relating these two aspects.

\section{Case Study Setup}
\label{sec:setup}

For the accomplishment of the case study was used as a set of software belonging to the Hibernate ecosystem to have their Technical Debt cataloged. For each software repository a Technical Debt Catalog of source code was constructed, tools were used in this process: SonarQube to identify the problems present in the source code and TD-Tracker for the cataloging of Technical Debt. Then, the analysis stage was carried out.

\subsection{Hibernate Ecosystem: Selecting Software Projects}
\label{sec:selection}

In the case study we used project software belonging to the Hibernate ecosystem\footnote{\url{https://github.com/hibernate/}}, not yet explored in the literature. To enable an analysis to be carried out, the software of this ecosystem must follow the following criteria that restrict our study:
The software should follow the Object-Oriented Paradigm and use the Java programming language in its development; The software project must be under development for a significant period, and may already have been finalized or still under development; The software must have several releases, since one of our analyzes aims to investigate the variation in the distribution  Code Debt rework time to each severity group over the development time.

The Hibernate ecosystem consists of 36 software repositories, however, it was necessary to disregard some of this software in our case study because they did not meet the criteria established for the scope of this work. Therefore, we selected 10 software repositories of the Hibernate ecosystem (Table~\ref{tab:projetos-software}) to compose our case study.

\begin{table}[!ht]
\centering
\caption{Software projects used in the case study}
\small
\begin{tabular}{lccc}
\hline
\textbf{Project} & \textbf{Developers} & \textbf{LOC} \\ \hline
Asciidoctor Extensions  &   2 &  292 \\ 
Commons Annotations     &  15 &   2K \\ 
HQL Parser              &  12 &   5K \\ 
Metamodelgen            &   8 &   6K \\ 
OGM                     &  55 & 111K \\ 
ORM                     & 461 & 748K \\ 
Search                  &  67 & 334K \\ 
Shards                  &   3 &  23K \\ 
Tools                   &  10 &  24K \\ 
Validator               &  77 &  98K \\ \hline
\end{tabular}
\label{tab:projetos-software}
\end{table}

\subsection{SonarQube}
\label{subsec:sonar}
The SonarQube\footnote{\url{https://www.sonarqube.org}}, is a code management open platform dedicated to analyzing and measuring the technical quality of source code. It also allows reporting a listing and information on potential \textit{issues}, to support continuous improvements in the quality of code. There are other tools available to identify problems in source code, however, SonarQube has the characteristic of detecting a large number of problems~\cite{digkas:2018}. Also, was used to identify code debt in other related works~\cite{jesus:2017}~\cite{digkas:2017}~\cite{digkas:2018}.








The source code inspection performed by SonarQube indicates to a \textit{Issues} whenever a part of the code breaks some coding rule. The \textit{Issues} are classified in the categories: Bugs, Vulnerabilities and Code Smells. Besides being associated to a degree of severity: \textit{\textbf{Blocker}}: Bug with high probability of impacting software behavior, must be corrected immediately; \textit{\textbf{Critical}}: Bug with low probability of impacting software behavior, code should be reviewed; \textit{\textbf{Major}}: Quality failure that may severely affect developer productivity; \textit{\textbf{Minor}}: Quality failure that may slightly affect developer productivity;

It is still estimated to each Issue a time of rework (in minutes), that is, the time needed to solve the problem and thus pay the Code Debt present in a certain part of the source code. Thus, the SonarQube tool was used in this work to perform the static analysis by inspecting the multiple versions of the source code of the software projects, identifying the issues characterized as Code Debt.

\subsection{TD-Tracker} 
\label{subsec:tracker}

TD-Tracker is a tool that supports the process of identification and mainly of cataloging Technical Debt. It makes use of an integrated catalog used to record the properties of each identified Technical Debt~\cite{foganholiTD:2015}~\cite{foganholi:2015}. TD-Tracker allows support for the cataloging of different types of Technical Debt, being they debts of architecture, code, test, documentation, and defects.

\begin{figure*}[!ht]
\centering\includegraphics[scale=0.72]{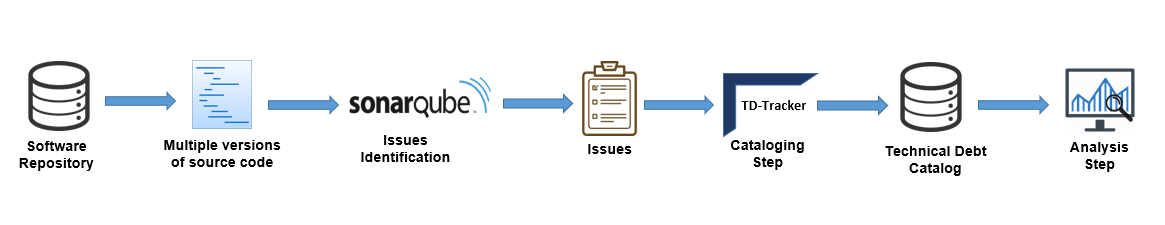}
\caption{Schematization of the process of cataloging Technical Debt}
\label{fig:esquema-metodologia}
\end{figure*}

TD-Tracker was used in conjunction with SonarQube to perform the cataloging of Technical Debt. While SonarQube performs the inspection in the source code and points out the existing problems, in turn, TD-Tracker catalogs these problems in the form of Code Debt. In Figure~\ref{fig:esquema-metodologia} is illustrated the scheme that characterizes the cataloging process.

\subsection{Technical Debt Catalog}
\label{sec:catalog}

A catalog was constructed for each Hibernate ecosystem software selected to be part of the case study, totaling ten Technical Debt Catalogs. Each Code Debt instance has a set of attributes that describe its characteristics. The following are each of the attributes of a Code Debt present in the catalogs:

\begin{itemize}

\item \textbf{Description}: Description of the problem that caused the Technical Debt.
\item \textbf{Type}: Describes the type of Technical Debt, in this case, only Code Debt is being studied.
\item \textbf{Project\_id}: Identifies the version code that the Technical Debt belongs to.
\item \textbf{Severity}: Identifies the severity of the problem that caused Code Debt, the greater the severity the greater the impact it can cause.
\item \textbf{Debt}: Identifies the time (in minutes) required for payment of Code Debt, time estimated by SonarQube can be configured, however for this work the standard settings of the SonarQube tool have been maintained.
\item \textbf{Location}: Identifies the path (package, class, and line) in the Code Debt is located within a version of the software.

\end{itemize}

Technical Debt Catalogs are one of the contributions of this work and can be used in other studies directed to the Technical Debt metaphor, such as: be used as a data set for research in Technical Debt Visualization, since it is reported in literature mappings~\cite{rios:2018}~\cite{alves:2016} that this is still an area that needs exploitation. The Technical Debt Catalogs are available at \url{[BLIND]}.


\section{Case Study: Results and Discussions}
\label{sec:resultados}

In this section, we present and discuss the results obtained by the case study, considering the research questions presented in Section~\ref{sec:introduction}.

\vspace{3 mm}
\noindent \fbox {
    \parbox{8.4cm}{RQ1: What is the distribution of rework time for each severity group of Code Debt? Is this distribution similar in all analyzed software of this ecosystem?}
}
\vspace{1 mm}


\begin{figure*}
  \centering
  
  \subfloat[\label{fig:rq1-a}]{%
  \includegraphics[width=0.48\textwidth]{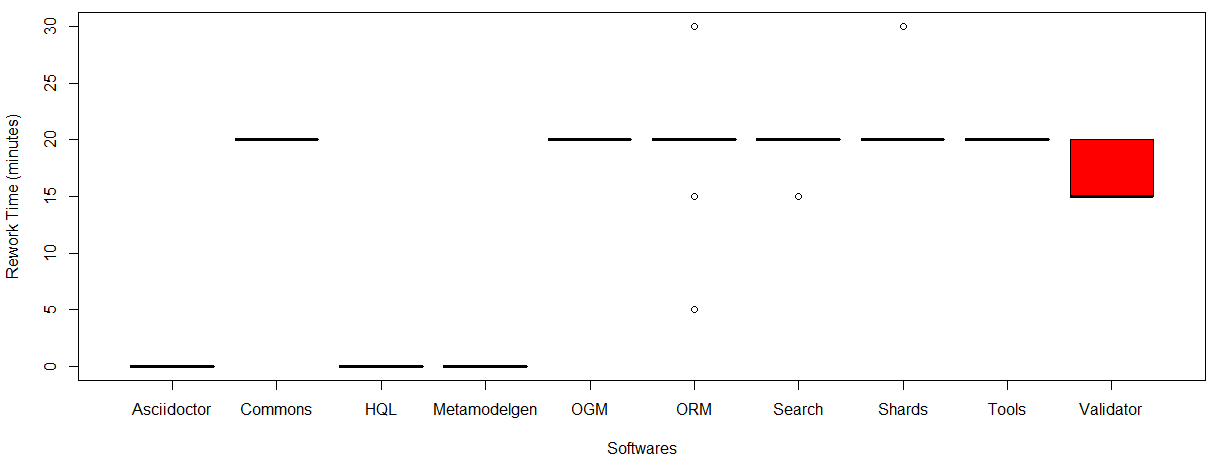}}
  \qquad
  \subfloat[\label{fig:rq1-b}]{%
  \includegraphics[width=0.48\textwidth]{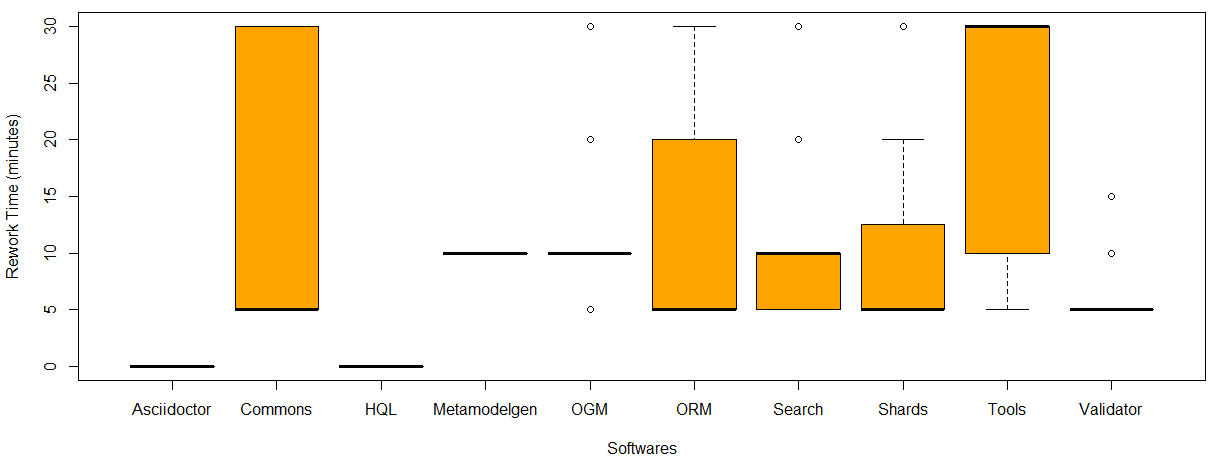}}
    \\
  
  \subfloat[\label{fig:rq1-c}]{%
  \includegraphics[width=0.48\textwidth]{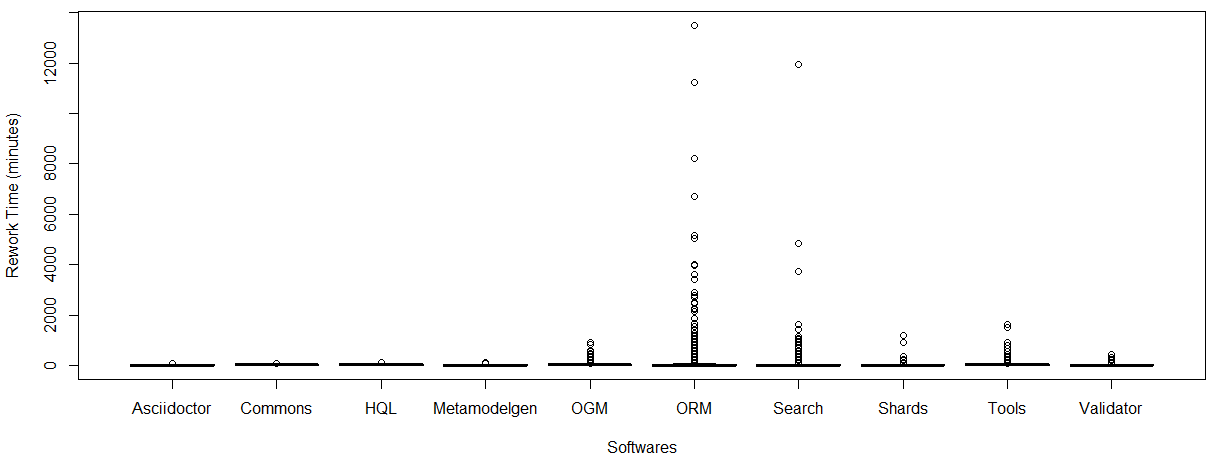}}
    \qquad
  \subfloat[\label{fig:rq1-d}]{%
  \includegraphics[width=0.48\textwidth]{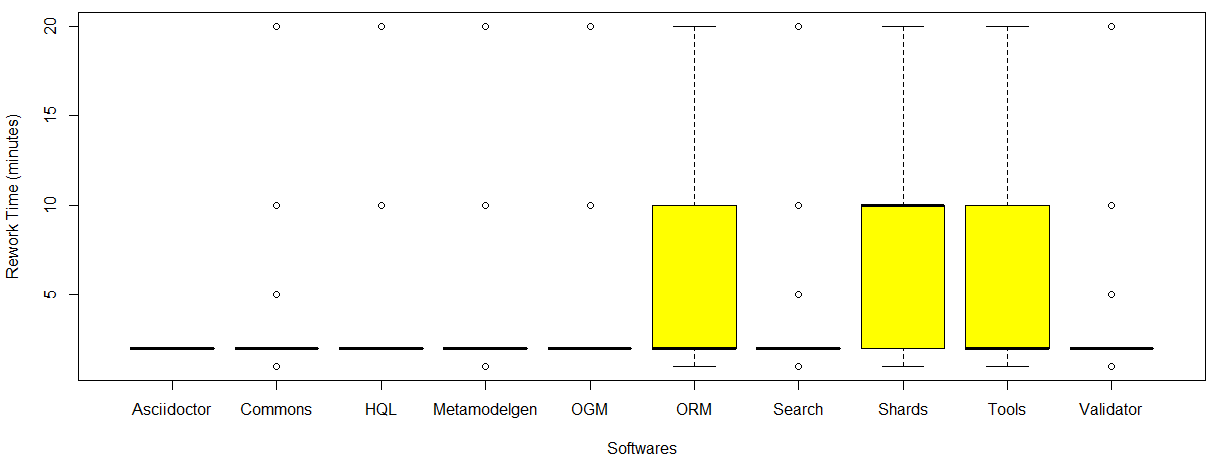}}
  
  \caption{Distribution of the rework time of each severity group: \textit{Blocker} (a), \textit{Critical} (b), \textit{Major} (c) and \textit{Minor} (d), of the Code Debt present in the last version of the Hibernate ecosystem softwares.}
  \label{fig:analise-rq1-1} 
\end{figure*}

The distribution of the rework time of each Code Debt severity group is presented in Figure~\ref{fig:analise-rq1-1}. Regarding the Code Debt with greatest impact to software functionalities -- \textit{Blocker} and \textit{Critical} severity groups --  (see Figure~\ref{fig:rq1-a}), three software do not present \textit{Blocker} severity, they are: Asciidoctor, HQL, Metamodelgen. Regarding other software, most \textit{Blocker} severity code debt have rework times of 20 minutes; we still observe outliers that indicate \textit{Blocker}, with rework time of 5, 15 and 30 minutes -- the minimum time to solve a Code Debt of \textit{Blocker} severity, in this ecosystem, is 5 minutes and maximum at 30 minutes. In the \textit{Critical} severity group -- Figure~\ref{fig:rq1-b} -- two software do not present such severity, they are Asciidoctor and HQL; regarding other software, the \textit{Critical} severity present rework time average between 5 and 10 minutes. Differently from the previous group, for \textit{Critical} severity there is a significant variation of rework time between different software analyzed. However, the minimum rework time is 5 minutes and the maximum is 30 minutes for this group.

\begin{figure}[!ht]
\centering\includegraphics[width=0.48\textwidth]{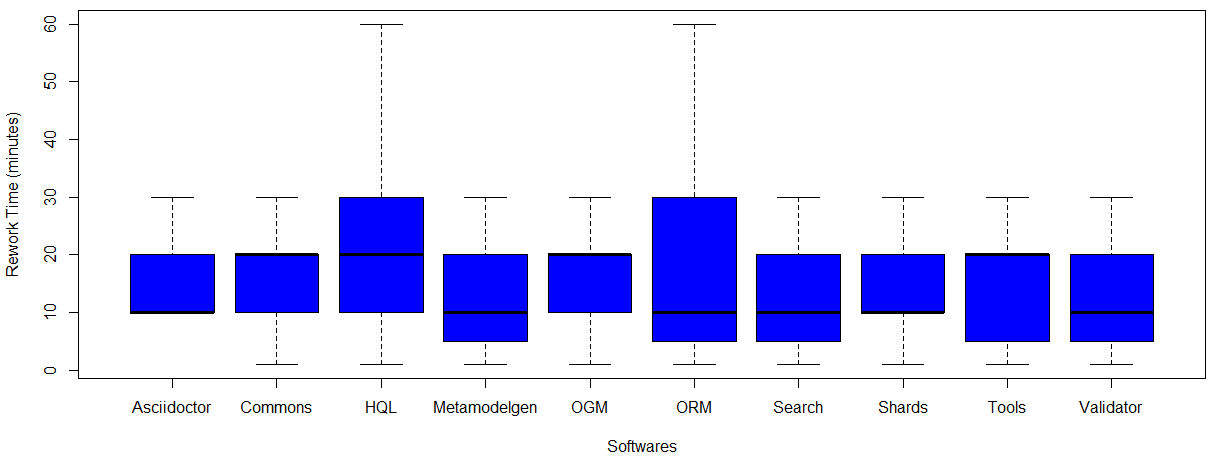}
\caption{Distribution of rework time of Code Debt with \textit{Major} severity present in the latest version of the Hibernate ecosystem softwares, disregarding outliers.}
\label{fig:analise-rq1-2}
\end{figure}

\textit{Major} and \textit{Minor} severity groups impacted negatively software maintenance capacity. One may observe that \textit{Major} severity group presents high rework times in comparison to the other severity groups -- Figure~\ref{fig:rq1-c}. We investigate the sort of Code Debt and they are mostly related to Code Duplication problems; the most duplications are in a code snippet, the longer the rework time is to solve this problem. Ignoring the outliers present in the severity group -- Figure~\ref{fig:analise-rq1-2}. 

\textit{Major} Code Debt present average rework times 10 and 20 minutes (discarding outliers), and minimum and maximum rework time 1 and 30 minutes, respectively, for most software. At last, the \textit{Minor} severity group has on average 2 minute rework times, a minimum of 1 minute and a maximum of 20 minutes, and in the Hibernate ecosystem, the Code Debt belonging to this group of severity have the lowest maximum rework time.

\textit{Major} Code Debt severity have high rework time, therefore, they will be hardly paid, considering the high effort required to perform this action and the low profit obtained. Solving those problems results in improved maintainability, however, the priority is to solve problems that directly affect the software functionality. Managers and Technical Debt management team, aware of problems that affect the capacity of maintenance and the high rework time, could perform training and guide developers to avoid problems, such as code duplication. 

Finally, we emphasize that Code Debt that impairs software functionality directly -- \textit{Blocker} and \textit{Critical} severity -- should be the first to be paid, since they have low rework times (maximum of 30 minutes identified) and would result in a higher quality gain for the software.

\vspace{3 mm}
\noindent \fbox {
    \parbox{8.4cm}{RQ2: Is the distribution of rework time for each severity group experiencing significant variation across the multiple versions of each software of the Hibernate ecosystem?}
}
\vspace{1 mm}


Analyzing each version of software projects belonging to the Hibernate ecosystem, we delineate the rework time distribution of each severity group and, thus, we observe the variation of this distribution along the development time of each software. 

In general, we observed that most analyzed software presents a little variation on rework time distribution over time. In general, the variations happen mainly at initial versions, and throughout the development, it tends to stabilize and maintain a similar distribution. This behavior happens because the initial versions undergo many modifications, generating a variation to rework time distribution of Code Debt. Over time, the acquired maturity in the development of each software, a smaller amount of modifications is necessary what results in the stabilization of rework time distribution. Therefore, all severity groups, in general, exhibit the same behavior: variations at initial versions or no variation in their distribution of rework time.

\vspace{3 mm}
\noindent \fbox {
    \parbox{8.4cm}{RQ3: Analyzing the distribution of the total amount of Code Debt of each severity group, which groups are most representative in the Hibernate ecosystem?}
}
\vspace{1 mm}

\begin{figure}[!ht]
\centering\includegraphics[scale=0.4]{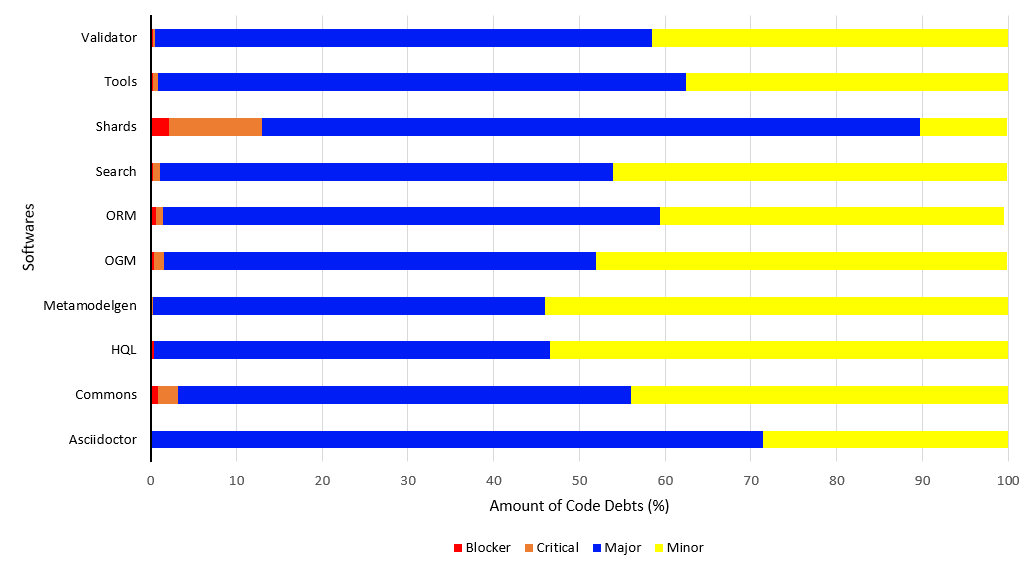}
\caption{Distribution of the total amount of Code Debt distributed by severity groups.}
\label{fig:analise-rq3}
\end{figure}

To answer this question, we verified the total amount of Code Debt of each severity group for each Hibernate ecosystem software. Then, we compared this distribution among all software to this ecosystem -- Figure~\ref{fig:analise-rq3}.

Analyzing the distribution of Code Debt of each severity group, we found that most of the Code Debt is related to problems that are harmful to software maintenance (\textit{Major} and \textit{Minor} severity), being the groups of the severity of greater representativeness in the Hibernate ecosystem. The positive aspect was the almost insignificant number of Code Debt of greater severity (\textit{Blocker} and \textit{Critical}), that directly impact the functionality of the software, only the Hibernate Shards and Hibernate Commons software present a significant amount of problems with such severities. This behavior is analogous in all versions of each of the analyzed software.

RQ1 has pointed out that \textit{Major} severity requires longer rework time to be solved in comparison to other severities. RQ3 pointed out the majority to \textit{Major} -- high number of occurrence (see Figure~\ref{fig:analise-rq3}). This Code Debt appears in large numbers in the multiple versions. The large amount and high rework time indicate difficulty to manually pay Code Debt. Thus, the use of automated refactoring techniques are necessary to pay \textit{Major} severity. The analysis of the effectiveness and possible techniques to be applied can be considered future work.

\vspace{3 mm}
\noindent \fbox {
    \parbox{8.4cm}{RQ4: Observing the evolution of the total rework time of each severity group, is the Code Debt controlled in the Hibernate Ecosystem?}
}
\vspace{1 mm}

The focus to this research question is the evolution of the total rework time required for the solution of all the Code Debt present in each analyzed software and, consequently, to verify if the Technical Debt related to each severity group is controlled, being paid or accumulated in this ecosystem. For that, considering each of the severity groups, the total rework time of each software version was obtained. This time was normalized between 0 and 1 to highlight not the amount but rather the evolution factor of the reworking time required to solve the problems of each severity group.

\begin{figure*}[!ht]
  \centering

  \subfloat[\label{fig:rq4-a1}]{%
  \includegraphics[width=0.33\textwidth]{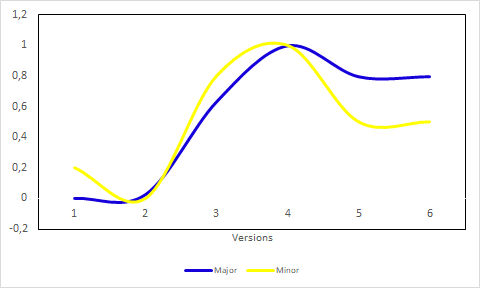}}
    \subfloat[\label{fig:rq4-b1}]{%
  \includegraphics[width=0.33\textwidth]{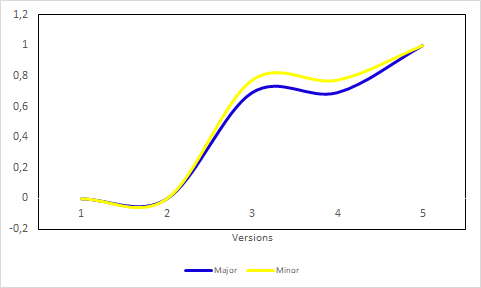}}
  \subfloat[\label{fig:rq4-c1}]{%
  \includegraphics[width=0.33\textwidth]{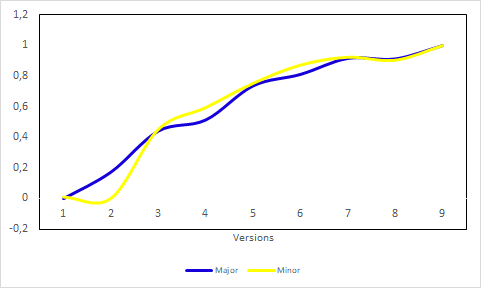}}
  
   \subfloat[\label{fig:rq4-d1}]{%
  \includegraphics[width=0.33\textwidth]{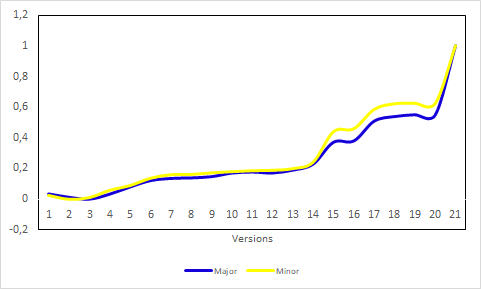}}
  \subfloat[\label{fig:rq4-e1}]{%
  \includegraphics[width=0.33\textwidth]{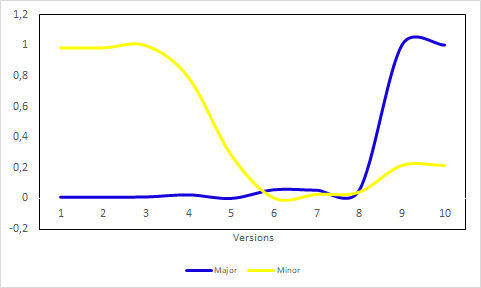}}
  \subfloat[\label{fig:rq4-f1}]{%
  \includegraphics[width=0.33\textwidth]{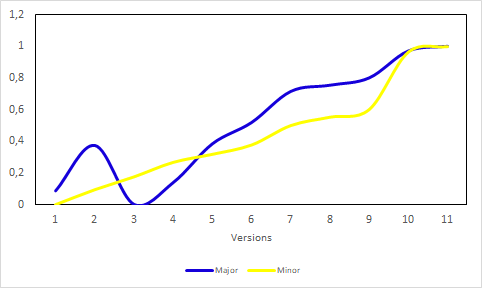}}  
  
  \caption{Rework time evolution to \textit{Major} and \textit{Minor} severity: HQL (a), Metamodelgen (b), OGM (c), Search (d), Tools (e) and Validator (f)}
  \label{fig:analise-rq4-1} 
\end{figure*}

\begin{figure*}[!ht]
  \centering

  \subfloat[\label{fig:rq4-a2}]{%
  \includegraphics[width=0.33\textwidth]{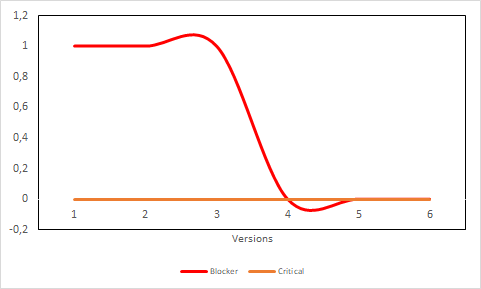}}
  \subfloat[\label{fig:rq4-b2}]{%
  \includegraphics[width=0.33\textwidth]{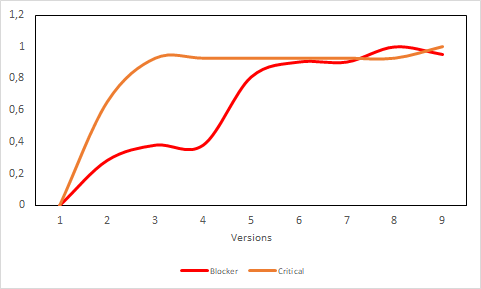}}
  \subfloat[\label{fig:rq4-c2}]{%
  \includegraphics[width=0.33\textwidth]{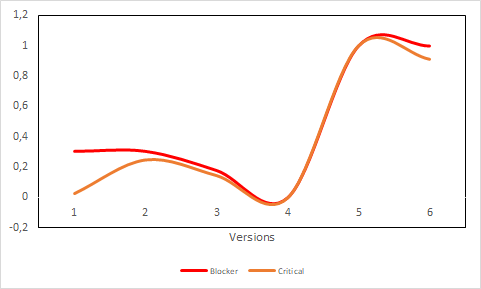}}
  	\\
  	
   \subfloat[\label{fig:rq4-d2}]{%
  \includegraphics[width=0.33\textwidth]{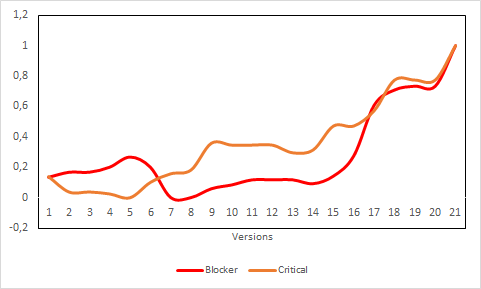}}
  \subfloat[\label{fig:rq4-e2}]{%
  \includegraphics[width=0.33\textwidth]{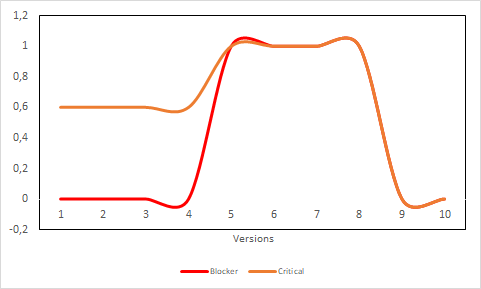}}
  \subfloat[\label{fig:rq4-f2}]{%
  \includegraphics[width=0.33\textwidth]{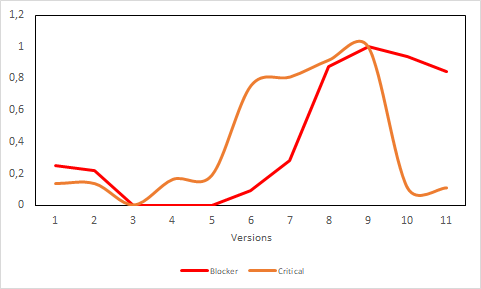}}  
  
  \caption{Rework time evolution to \textit{Blocker} and \textit{Critical} severity: Commons (a), OGM (b), ORM (c), Shards (d), Tools (e) and Validator (f)}
  \label{fig:analise-rq4-2} 
\end{figure*}

The Figures~\ref{fig:analise-rq4-1} e~\ref{fig:analise-rq4-2} illustrate, respectively, the evolution of some software of the Hibernate ecosystem in perspective of the total rework time of Code Debt detrimental to maintenance (\textit{Major} and \textit{Minor} severity) and Code Debt that directly affect the operation of the software (\textit{Blocker} and \textit{Critical} severity).

In general, observing the evolution of the Code Debt rework time of \textit{Major} and \textit{Minor} severity,  the increase of rework time was observed during the software development. To more than 50\% of the analyzed software, the rework time of both severities tends to increase. This leads one to believe that they are not controlled by the development team, being generated and not paid over the development period. The minority of the ecosystem software has presented decreasing rework time of one severity, they are Hibernate Commons, Shards and Tools.

Regarding the rework time evolution of \textit{Blocker} and \textit{Critical} severity Code Debt, both harmful to the software operation, we observed a stabilized evolution to the minority of analyzed software, with low and constant rework times to all software versions (Hibernate Asciidoctor, Commons and Metamodelgen). Other software presents an evolution characteristic followed by decreasing points. This situation may indicate that problems with this severity are being controlled and solved.

\section{Final Remarks}
\label{sec:conclusoes}

The Technical Debt accumulation results in increased cost of the maintenance phase and a decrease in software quality. Knowing the behavior of debt can support to create efficient strategies in activities of prevention or payment of Technical Debt. We describe a case study carried out with the Hibernate ecosystem, analyze the behavior of Code Debt with the focus on aspects of severity and rework time. Technical Debt Catalogs were built which were later analyzed.

The results of the study revealed that most Code Debt is detrimental to code maintainability, \textit{Major} and \textit{Minor} severity, and in particular, those of \textit{Major} severity are those that need more time to be solved. We also investigated the variation of the rework time distribution of the Code Debt of each severity group and found that when there is variation, in general, it occurs in the early versions of development.

Technical Debt Catalogs are positive contributions of this study since they can be used in other works involving Technical Debt. As future work we also intend to use the catalogs as a database, to investigate the possible benefits that visualization techniques can provide to support the monitoring and management of Technical Debt, since this is a lacuna pointed out in the literature, requiring further investigation~\cite{rios:2018}~\cite{alves:2016}.

\section*{Acknowledgement}
\label{sec:acknowledgement}

\noindent We are grateful to [BLIND] Research Agency for financial support to development of this work.



\bibliographystyle{IEEEtran}
\bibliography{bibtex}
\end{document}